\documentstyle[prb,aps,epsfig,multicol]{revtex}

\begin{document}
\title{Electronic Structure and Heavy Fermion Behavior in LiV$_{2}$O$_{4}$}
\author{D.J. Singh}
\address{Code 6391, Naval Research Laboratory, Washington, DC 20375}
\author{P. Blaha and K. Schwarz}
\address{Institut f\"ur Physik und
Theoretische Cheme, TU Wien, A-1060 Wien, Austria}
\author{I.I. Mazin}
\address{Code 6391, Naval Research Laboratory, Washington, DC 20375}
\date{\today}
\maketitle

\begin{abstract}
First principles density functional calculations
of the electronic and magnetic properties
of spinel-structure LiV$_{2}$O$_{4}$ have been
performed using the full potential linearized augmented planewave method.
The calculations show that the electronic structure near the Fermi
energy consists of a manifold of 12 bands derived from V $t_{2g}$ states,
weakly hybridized with O p states.
While the total width of this active manifold is approximately
2 eV, it may be roughly decomposed into two groups: high velocity bands and
flatter bands, although these mix in density functional calculations.
The flat bands, which are the more atomic-like lead to a high density
of states and magnetic instability of local moment character.
The value of the on-site exchange energy is sensitive to the exact
exchange correlation parameterization used in the calculations,
but is much larger than
the interaction between neighboring spins,
reflecting the weak coupling of the magnetic system with the
high velocity bands.
A scenario for the observed
heavy fermion behavior is discussed in which
conduction electrons in the dispersive
bands are weakly scattered by local moments
associated with strongly correlated electrons
in the heavy bands.
This is analogous to that in conventional Kondo
type heavy fermions, but is unusual in that both the local moments
and conduction electrons come from the same d-manifold.
\end{abstract}
\begin{multicols}{2}

\section{Introduction}

Heavy fermion (HF) materials are typically inter-metallic compounds
containing Ce, U or Yb atoms.
They are characterized by the usual Landau Fermi liquid scaling properties,
but only at very low temperature (beginning
as low as 0.3 K, depending on the material)
and with extraordinarily strongly renormalized effective masses,
$m^{*} \approx 100 - 1000 m_e$. \cite{anders,stewart,aeppli}
They show apparent local moment paramagnetic behavior with strongly
increasing spin susceptibility, $\chi$,
and specific heat coefficient $\gamma$ with decreasing temperature,
but do not order magnetically and eventually settle at low $T$
into a state with constant $\gamma$ and $\chi$ and
Wilson ratio near unity.
Following extensive investigation over many years,
a basic understanding of the
phenomena has been established. The origin is a many body effect associated
with the interaction of itinerant conduction electrons with strongly
correlated f-electrons on local                            
moment rare earth ions. Thus the discovery by Kondo and co-workers of
HF behavior \cite{kondo1} with $\gamma \approx 0.42$ J/mol K$^2$,
in the transition metal oxide LiV$_{2}$O$_{4}$
was both remarkable and unexpected as the material does not appear to
fit into this framework.

As may be expected, this discovery has led to a series of detailed
experiments that have confirmed the original result with properties
characteristic of the heaviest f-electron HF
materials, \cite{chmaissem,mahajan,fujiwara,kondo2,johnston,krimmel}
although it should be noted that Fujiwara and co-workers proposed an alternate
explanation of their NMR data within a spin-fluctuation
framework. \cite{fujiwara}

Two first principles band structure calculations have recently
appeared \cite{eyert,anisimov}, though both use
spherical approximations for the potential. These
reasonably agree with each other but not as well with the
full potential calculations presented here.
Their conclusions are, however, different: Eyert
et al. \cite{eyert} suggest that LiV$_{2}$O$_{4}$ is not a
``true'' HF material in the sense that they ascribe the specific heat
enhancement to spin fluctuations arising from frustrated
antiferromagnetism on the V sublattice;
Anisimov et al. \cite{anisimov} advocate a
separation of V d electrons into two sets, a localized and an 
itinerant one, playing the role of the f and conduction
electrons in conventional HF materials.
A related model is discussed by Varma. \cite{varma}
We argue that the true HF picture is likely, since it is compatible with 
the electronic structure obtained and is consistent with the experimental 
situation.
 
\begin{figure}[tbp]
\centerline{\epsfig{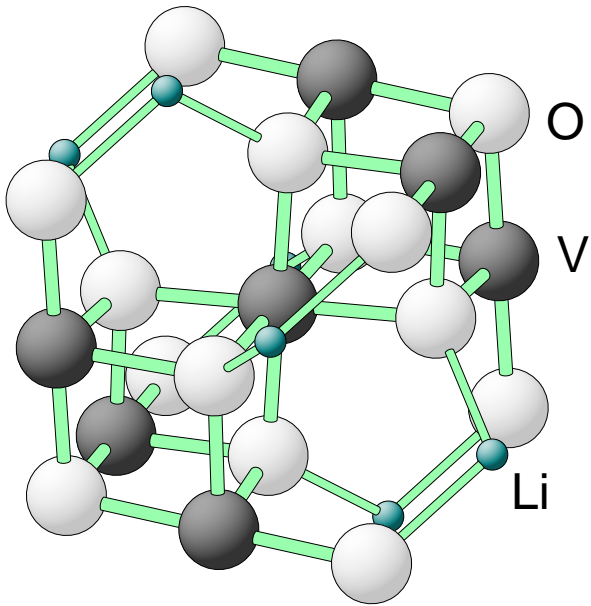}}
\vspace{0.15in}
\setlength{\columnwidth}{3.2in} \nopagebreak
\caption{
Structure of LiV$_2$O$_4$ showing the local coordination of a V atom.
Note that central V in the figure
is at the center of an O octahedron oriented along
the 100 axes and on the joint vertex of two corner sharing V tetrahedra,
leading to the rhombohedral site symmetry.}
\label{struct}
\end{figure}

LiV$_{2}$O$_{4}$ occurs in an undistorted cubic fcc spinel structure as
shown in Fig. \ref{struct}.
The V$^{+3.5}$ (3$d$ 1.5) ions are located in very slightly slightly distorted
O octahedra.
These V ions form corner sharing tetrahedra ordered on the fcc lattice such that
each V has six neighbors of the same kind. Thus the V sublattice can be
viewed as an fcc lattice with every second atom removed. Alternately,
it can be viewed as a collection of chains running along all the (110) type
directions with each V participating in three chains.
This structure leads to strong
geometric frustration of antiferromagnetic (AF) interactions,
although of course ferromagnetic (FM) interactions are unfrustrated.
As noted, LiV$_2$O$_4$ is paramagnetic at low temperature.
However, Krimmel and co-workers \cite{krimmel} have
reported quasi-elastic neutron scattering measurements
that show predominantly FM
fluctuations above 40K with substantial observable AFM fluctuations
only at lower temperatures,
where they significantly contribute to the spectra.

\section{Band Structure and Magnetism}

In this structural configuration, we find that
the O 2p bands are below and well separated from the transition
metal derived bands so that both the Li-O and V-O bonds are strongly
ionic. The higher lying 3$d$ derived orbitals are crystal field
split into a 6-fold degenerate (3 per spin) $t_{2g}$ manifold and
a higher lying 4-fold degenerate $e_g$ manifold.
Neither the crystal structure nor the electronic configuration is
particularly favorable for substantial V 3$d$--O 2$p$ hybridization,
so narrow d bands are obtained with clean crystal field
gaps between the $t_{2g}$ and $e_g$ manifolds.
We do not get noticeable mixing of these manifolds.
Thus the electronic structure relevant to low energy excitations
may be described as a
manifold of 12 narrow $t_{2g}$ bands $1/4$ filled with 6 electrons
(note that there are 4 V ions per
cell) with no other nearby bands. Much of the bandwidth is
due to direct V-V hopping.
At first glance it is hard to recognize the connection
between this electronic structure and that of the f-electron HF compounds.

We find, using density functional calculations with the
full potential linearized
augmented planewave (LAPW) method, \cite{singhlapw,lapwnote,blahalapw,weilapw}
that the $t_{2g}$ manifold in fact contains two
different types of carriers as was also emphasized by
Anisimov et al. \cite{anisimov}.
Calculations with the tight-binding linear muffin tin orbitals (TB-LMTO)
method \cite{OKA75}
were used to analyze the band symmetries.
One carrier type is
analogous to the itinerant conduction electrons of conventional HF
materials and the others can play the role of the
strongly correlated local moment f-electrons.
This provides a way of obtaining conventional HF
behavior in LiV$_{2}$O$_{4}$ albeit in an unconventional way, at least as
regards the nature of the bare states.

\begin{figure}[tbp]
\centerline{\epsfig{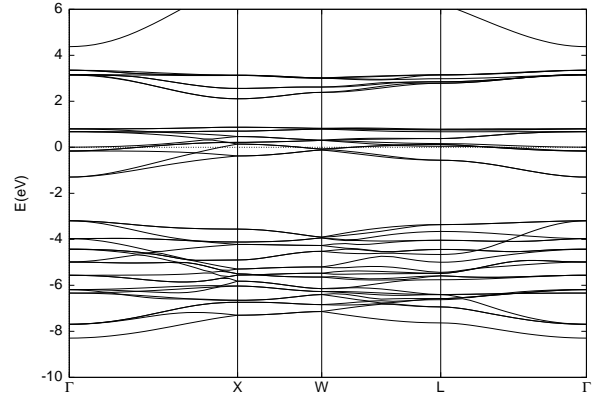}}
\vspace{0.15in} \setlength{\columnwidth}{3.2in} \caption{
Calculated band structure of non-spin-polarized LiV$_2$O$_4$.
The Fermi energy is denoted by the dashed horizontal line at 0.}
\label{nsp-bands} \end{figure}

\begin{figure}[tbp]
\centerline{\epsfig{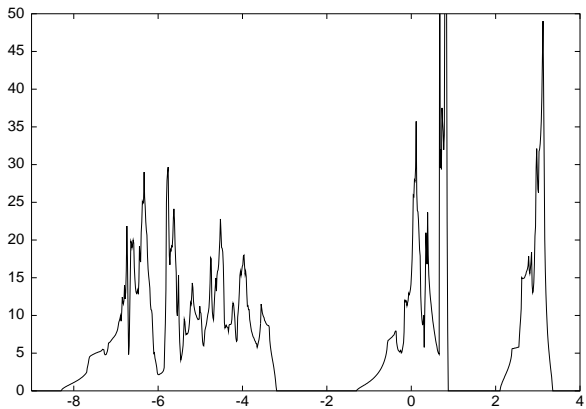}}
\vspace{0.15in}
\setlength{\columnwidth}{3.2in} 
\caption{
Electronic density of states of non-spin -polarized LiV$_2$O$_4$ in states
per eV per unit cell (4 V atoms) as a function of energy in eV. The Fermi
energy is at 0.}
\label{dos} \end{figure}

\begin{figure}[tbp]
\centerline{\epsfig{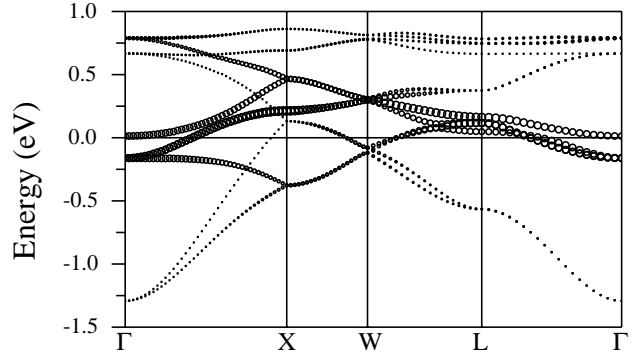}}
\vspace{0.15in}
\setlength{\columnwidth}{3.2in} \nopagebreak
\caption{
Calculated band structure of non -spin -polarized LiV$_2$O$_4$
showing the $t_{2g}$ manifold. The sizes of the points denote the 
calculated relative amounts of $a_{1g}$ character in the states.
The Fermi energy is denoted by
the dashed horizontal line at 0.}
\label{bands-small}
\end{figure}

The calculated local spin density approximation (LSDA)
band structure for non-spin-polarized (NSP)
LiV$_{2}$O$_{4}$ is shown in Fig. \ref{nsp-bands}.
The corresponding electronic density of states is given in Fig. \ref{dos}.
The O 2p bands span the range from
approximately $-8$ to $-3$ eV (relative to $E_F$).
The 12 $t_{2g}$ bands are separated from
these by a gap and have a width of slightly less than 2.5 eV.
The $e_g$ manifold is quite
narrow, characteristic of the bonding topology of spinels,
which features bent V-O-V bonds,
unfavorable for band formation via $e_g - p \sigma$ hybridization.
These are again well separated
from the active $t_{2g}$ bands by a gap of slightly less than 1.5 eV.
The dispersion of the $t_{2g}$ bands is in fact mostly derived from 
V-V hopping as was verified by removing the O orbitals from TB-LMTO
calculations.
A blow-up of the band structure showing the $t_{2g}$ bands, which
dominate the low energy physics, is given in Fig. \ref{bands-small}.
Although a high on-site Coulomb repulsion, $U$,
is not expected in early 3$d$ compounds like this, significant
correlation effects on this band structure should not be
excluded since the active manifold is narrow.
Photoemission measurements \cite{fujimori} imply $U \approx 2$ eV, which is
comparable to the bandwidth, and therefore consistent with at least
moderate correlation effects in this multi-band system.

Spin polarized calculations, both within the LSDA and with
generalized gradient approximations (GGA) \cite{perdew}
show a considerable magnetic instability, of strong local
moment character. Calculations were done for ferromagnetic,
antiferromagnetic (AF --- 2 of 4 spins flipped per unit cell)
and ferri-magnetic (FiM --- 1 of 4 spins flipped) configurations.
Very low Hellman-Feynman forces on the symmetry unconstrained O
coordinates were obtained in the experimental crystal structure,
used in these calculations, independent of the magnetic order,
supporting the experimental structure
and implying only weak magneto-elastic effects
as may have been anticipated based on the weakly hybridized electronic
structure. Within the LSDA, a FM spin polarization of 1.15 $\mu_B$/V
with an energy 60 meV/V below the NSP energy is obtained. The AF and FiM
energies were both 5 meV/V lower than the FM. This is qualitatively
different from the results of Eyert $et\ al.$\cite{eyert}, who found
energies and moments closer to itinerant magnetic system consistent
with strong scattering of carriers by spin fluctuations at the LSDA level.

The GGA results are qualitatively similar to
the LSDA, but are somewhat more magnetic.
In the GGA the FM spin polarization is increased to 1.4 $\mu_B$/V,
which is a greater than usual sensitivity to the exchange correlation
functional. The magnetic properties are also unusually sensitive
to other details of the calculations: for instance, the atomic sphere
approximation (ASA), used in Refs.\onlinecite{eyert,anisimov},
as well as in our
LMTO calculations, leads to substantial underestimate of the magnetic
moment, namely 0.5  $\mu_B$/V. This is also unusual and calls for
caution when using ASA calculations for quantitative estimates
of the electronic parameters (cf. Ref.\onlinecite{anisimov}).
This is related to the physics discussed below.
First of all, the magnetization
comes almost entirely from one $d$ sub-manifold, resulting in a highly
non-spherical exchange potential, which is spherically averaged in
non-full-potential calculations reducing its effect.
Secondly, the NSP energy bands differ such that the full-potential band
structure has higher values of the density of states around the Fermi level,
and therefore a stronger Stoner instability. To test this,
extended Stoner calculations were
performed with the same Stoner $I$ (from the TB-LMTO),
but the LSDA TB-LMTO and LAPW densities of
states. This gave a FM magnetization of slightly more than 0.5 and 
slightly more than 1 $\mu_B$/V respectively, indicating that the changes in
the bare NSP bands are the more important factor, although it should be
noted that with the full exchange splitting the bands are non-rigid so
the extended Stoner is not fully valid and the result should be regarded
as qualitative.

\begin{figure}
\centerline{\epsfig{file=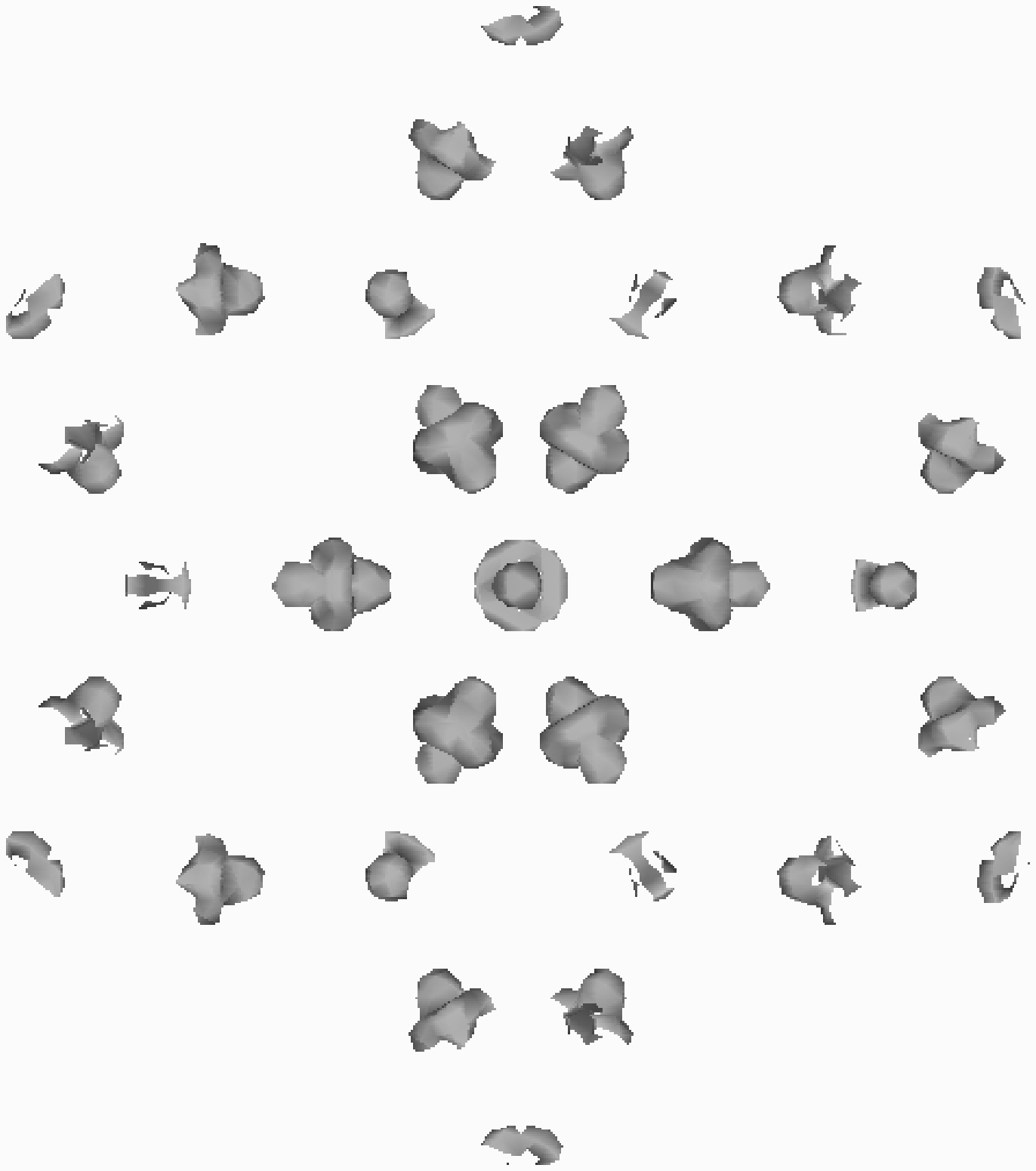,angle=0,width=2.50in}}
\centerline{\epsfig{file=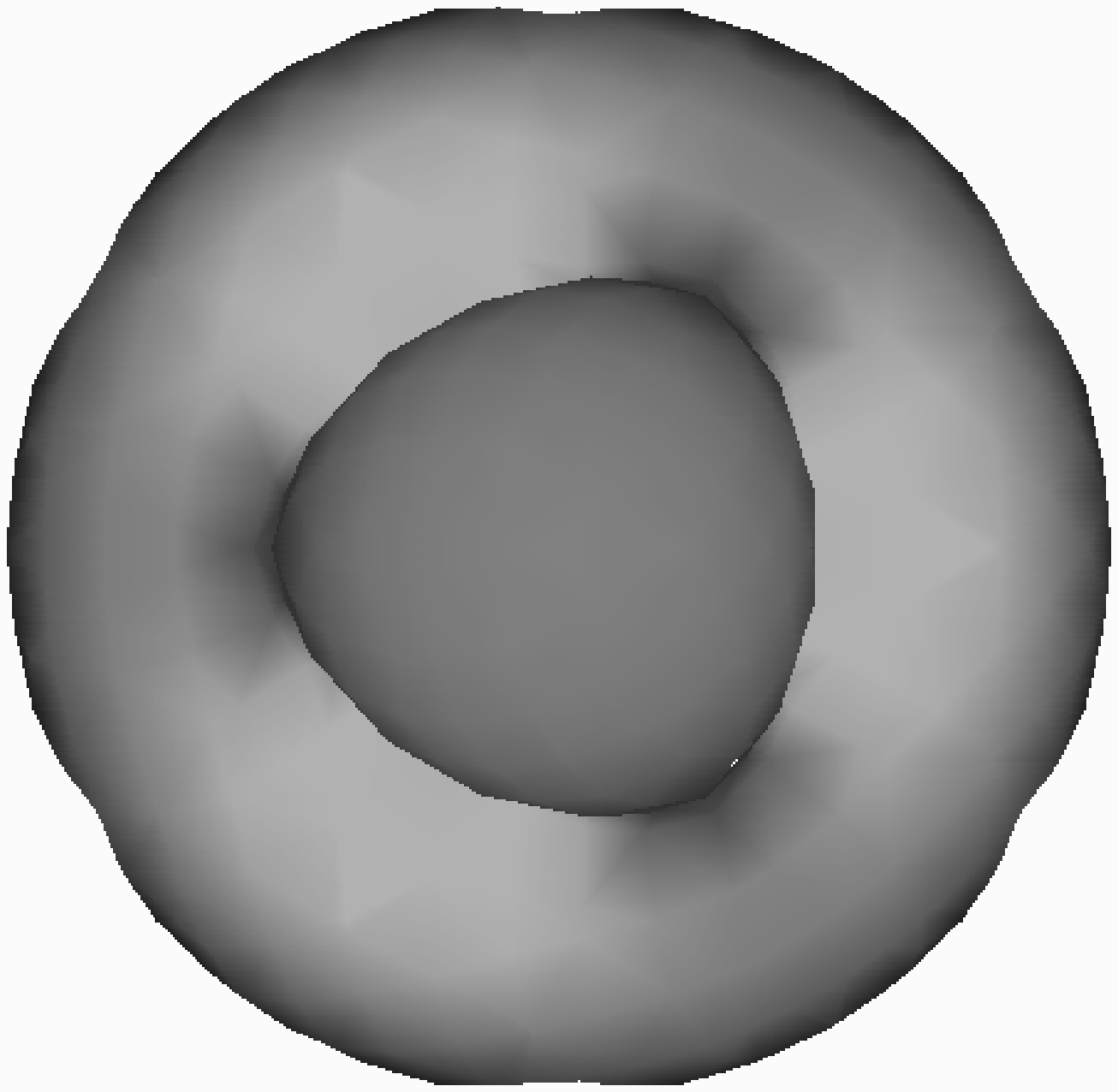,angle=0,width=2.50in}}
\centerline{\epsfig{file=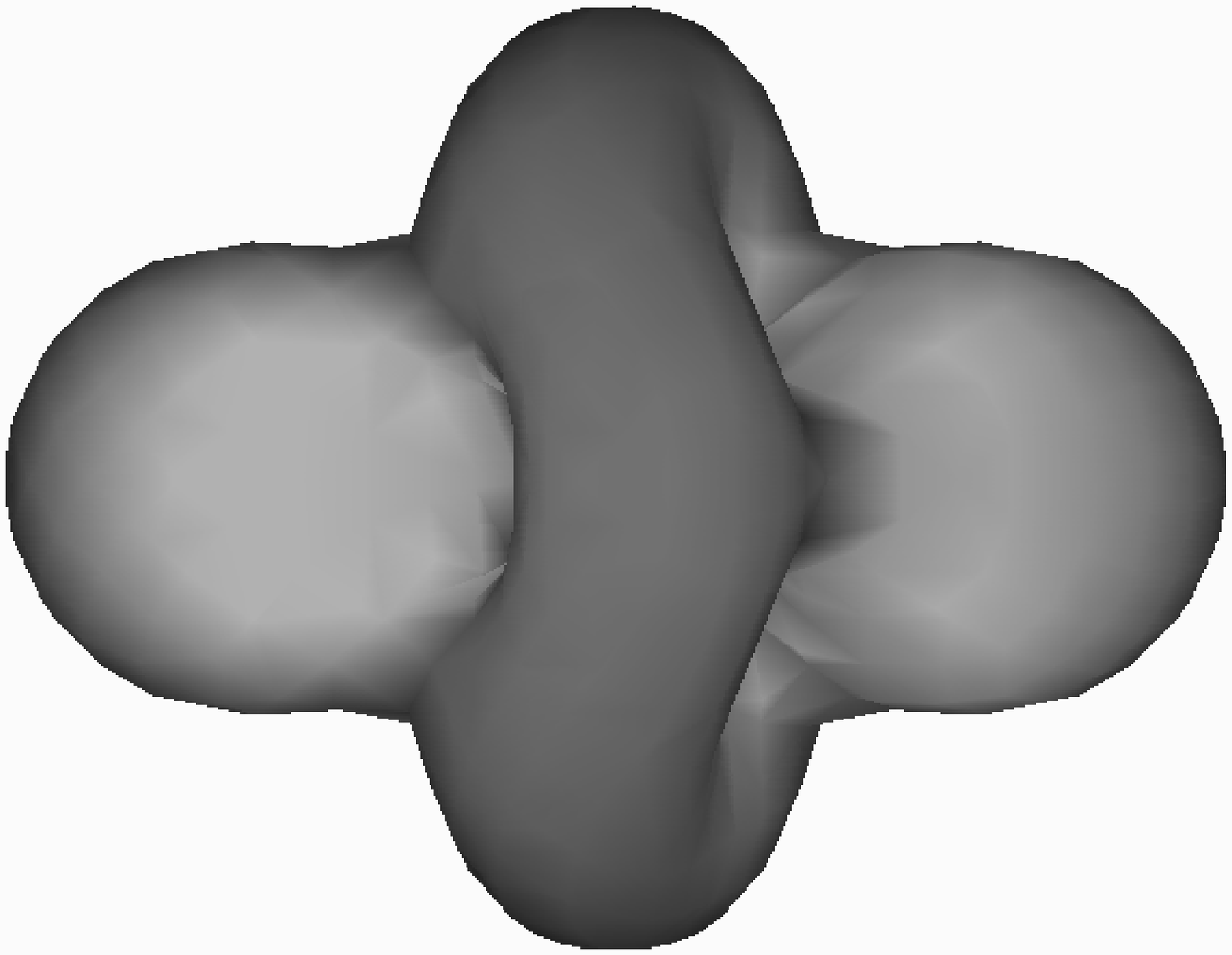,angle=0,width=2.50in}}
\vspace{0.15in}
\setlength{\columnwidth}{3.2in} \nopagebreak
\caption{
Calculated spin density of LiV$_2$O$_4$ with a ferromagnetic ordering.
The top panel shows
an 0.1 $\mu_B / a.u.^3$ isosurface viewed along a 111 direction, the middle
is the same view blown up around a single V ion, while the bottom panel is
the same on a -110. Note the predominant but not pure
$a_{1g}$ character. The small
$e_g^{\prime }$ admixture is responsible for the triangular rather than full
symmetry viewed from the top (middle panel)
and the buckling of the central "doughnut"
in the side view (bottom panel).}
\label{spindens}
\end{figure}

The calculated spin density (Fig. \ref{spindens}) is remarkable,
in that despite the nearly cubic environment, it strongly reflects the
weak rhombohedral symmetry. This is a result of the fact that
the hopping that gives rise to the $t_{2g}$ dispersions is largely V-V
and that the V neighbors are a rhombohedral environment.
The primary pseudocubic quantization axis is along the Cartesian 100 directions
because of the octahedral mostly O derived crystal field that splits the
$t_{2g}$ and $e_g$ manifolds. In the rhombohedral site symmetry there
is a further quantization of the $t_{2g}$ space based on the rhombohedral
111 axis, into a double degenerate $e_g$ space
and a single degenerate $a_{1g}$ manifold. In order to distinguish this
$e_g$ from the higher lying $e_g$ manifold split off by the octahedral
crystal field, we will denote it $e_g^{\prime }$ in the following.
The $e_g^{\prime }$ and $a_{1g}$ are at the same energy in LiV$_2$O$_4$ but are
locally orthogonal spaces.
The remarkable result, evident in Fig. \ref{spindens}
is that the magnetization is
derived almost entirely from the $a_{1g}$ sub-space.

We obtain almost complete polarization of the relevant $a_{1g}$
sub-manifold in our full-potential
LSDA and GGA calculations. Besides the higher resulting magnetic moments,
we obtain much clearer strong local moment behavior in these calculations.
The V moments as characterized by integrated spin magnetization
in the V LAPW spheres for the AF and FiM cases are within 9 percent
of the FM value.
The FiM and AF cases were energetically degenerate to within the
accuracy of the calculations at 5 meV/V below the
FM energy, consistent with weak AF exchange interactions
and magnetism that, as noted above, is very strongly local moment in character.

As mentioned, the dispersion of the $t_{2g}$ bands is
nearly exclusively derived from the V-V interaction. This includes the
nearest neighbor $dd\sigma $ and $dd\pi $ hoppings. Inverting the TB-LMTO
Hamiltonian we get a rough estimate, $t_{dd\sigma }/t_{dd\pi }\sim 3.5$ (and 
$t_{dd\delta }\ll t_{dd\pi }$ ). The 12$\times 12$ TB Hamiltonian in this
case is defined by the overlap integrals for each V-V bond, {\it e.g., }for
the (110) bond $t_{xy-xy}=\frac{3}{4}t_{dd\sigma },$ $t_{yz-xz}=t_{dd\pi }, $
$t_{xy-xz}=0$. An interesting property of this Hamiltonian is that when
$t_{dd\sigma }=\frac{8}{3}t_{dd\pi }$, two bands are flat.
 These bands are nonbonding combinations of $e_{g}^{\prime }$ states of the
 four V ions.
In
actual calculations this condition is satisfied only very approximately
(within $\approx$ 40 percent), but the corresponding states are still
very flat (the width is 0.2 eV) and located about 0.8 eV above the Fermi
level. Full-potential LAPW calculations reveal also two more relatively
flat bands of
the $e_{g}^{\prime }$
character right below these, but still
about 0.7 eV above the Fermi level.
On the other hand, a closer look reveals four more heavy bands 
(width $\approx$ 0.5 eV) that are located around the Fermi
level. Those have predominantly $a_{1g}$ character
and give rise to the high concentration of $a_{1g}$ character
around $E_F$ and the high density of states in the same region with
the resulting magnetic instability (cf. Fig. \ref{bands-small}).
The remaining bands are mixed in the LSDA;
$e_{g}^{\prime }$ character dominates and they have greater dispersion.

\section{Scenario for Heavy Fermion Behavior}

The variation of the effective mass is substantial: near the bottom of the
$t_{2g}$ manifold it is close to the free electron mass (which is unusually
small for $d$-bands), while near the Fermi level it is 4--5. From the
transport point of view, one can speak about two groups of carriers:
four light bands 
(comparable with the $sp$ bands in transition metals), and
four heavier bands 
(with their width of $\approx 0.5$ eV).
Additionally, there are four flat bands at about 0.7--0.8 eV above $E_F$.
The bands providing the heavy carriers like the light bands (width 2 eV)
cross the Fermi level but are
responsible for the calculated local moment magnetic instability.
These heavy bands near $E_F$
are dominated by $a_{1g}$ character, while the remainder of the
bands, although mixed are dominated by $e_{g}^{\prime }$ character.

In the scenario proposed here, the states relevant to the transport
properties are from the heavy and light bands near $E_F$;
these may constitute the
two classes of carriers mentioned above.
This connects with conventional
HF materials, in that LSDA calculations for these materials also
show band structures consisting of broader conduction bands and narrow
($W \ll U$)
bands at $E_F$ associated with local moments. In conventional HF materials
these two classes of bands are coupled, but weakly. The first evidence for
this in LiV$_2$O$_4$
is through the small (but no doubt overestimated in LSDA) inter-atomic
exchange interactions as seen by the near degeneracy of the various magnetic
configurations studied and the fact that although
the ferromagnetic band structure has exchange splittings in all bands,
very little polarization of the $e_{g}^{\prime }$
orbitals is induced as discussed 
above. Further, more clear signatures emerge from the transport functions,
particularly
$N(E_F)\langle v_F^2\rangle \propto  V \omega_p^2  \propto  V (n/m)$
where $v_F$ is the band velocity on the
Fermi surface, $V$ is the unit cell volume, $\omega_p$ is the
plasma frequency, $n/m$ is the optical $n/m$ and the average $\langle \rangle
$ is
over the Fermi surface. This function is dominated by more dispersive
bands due to the $\langle v_F^2\rangle$
factor, so that changes with magnetic order
are reflective of the interaction between the low dispersion states
responsible for the magnetism and the more dispersive conduction $a_{1g}$
derived bands. We obtain
$N(E_F)$ of 7.26 eV$^{-1}$ on a per formula unit basis (2 V atoms) for 
the paramagnetic calculation, and 7.23 eV$^{-1}$
for the FM majority spin channel,
1.58 eV$^{-1}$ for the FM minority channel,
2.96 eV$^{-1}$ for the FiM majority spin channel and
7.24 eV$^{-1}$ for the FiM minority channel.
The corresponding Fermi velocities (cm/s) are
$0.77\times 10^7$, $0.30 \times 10^7$, $1.37 \times 10^7$,
$0.60 \times 10^7$ and $0.75 \times 10^7$.
The LSDA values of $N(E_F)\langle  v_F^2\rangle$ stay within 20 percent of the
paramagnetic value.

So far all our results emerge from first principles density functional
calculations primarily within the LSDA using the full potential LAPW method.
This includes the separation into
two different types of carriers, $a_{1g}$ like and $e_g^{\prime}$ like
and the strongly spin polarized $a_{1g}$ like local moment magnetism.
Coulomb correlations as included approximately in the LSDA+U method
are not required. This is of some significance because in the LMTO
calculations of Ref. \onlinecite{anisimov} a $U$ of 3 eV was used to
obtain these, and this $U$
is about 50 percent greater than the value implied by photoemission and
which is also a value that would split the $e_g^{\prime}$ space into
strongly correlated upper and lower Hubbard bands. On the other hand,
with the screened $U$
less than or equal to the value indicated by photoemission,
as is normal, i.e. $U \alt 2$ eV, the light $e_g^{\prime}$ derived
bands would be weakly to moderately correlated and metallic as in conventional
HF while only the heavy $a_{1g}$ bands would be strongly correlated. In fact,
the actual situation may be more complicated because the two relevant
types of carriers are derived from the same $d$ manifold and therefore
may be more strongly coupled than in conventional HF, as discussed below.

\section{Discussion}

Even in weakly correlated materials, LSDA calculations
overestimate the extent of hybridization between different orbitals.
Thus one may anticipate the following artifacts in the LSDA calculations:
(1) the heavy $a_{1g}$ bands could be narrower than calculated and
(2) the mixture between $a_{1g}$ and $e_g^{\prime }$ bands could be weaker than
calculated and (3) as a result
the dominance of $a_{1g}$ character in the spin density
could be stronger than that obtained from the LSDA.
This is like the corresponding artifacts in band calculations
for f-band metals where the hybridization of the f-bands with
the conduction bands is overestimated in the LSDA.
These are partially local effects related to LSDA errors in the
description of the bonding
and not directly to Hubbard type correlations controlled by $U/W$
(the same effects are present in wide
band materials, where, for example, LSDA estimates of interatomic magnetic
couplings, $J$ are often overestimated). 
Additionally, the effect of
Hubbard type correlations, which are neglected in band structure calculations,
may be expected to be strong for the heavy mass carriers as 
taking $U \alt 2$ eV, as indicated by photoemission measurements
\cite{fujimori}, yields $U/W \alt 4$. At such values of $U/W$ one 
may expect a substantial suppression of charge fluctuations involving
the heavy mass bands along the lines of the normal f-electron HF
systems. However, the real situation is undoubtably
considerably more complex,
as some Hubbard correlation effects may be present in the bare light carrier
bands where $U/W \alt 1$
and exceptionally strong off-diagonal Hubbard effects may be expected as
well since the relevant bands (heavy and light) are both derived
from the V $t_{2g}$ space. It will be quite interesting to construct and
study many body models of this type in the HF regime.
Only the $t_{2g}$ orbitals and the interactions
among them should be needed.
There should be four light bands and four related flat bands away from $E_F$
per unit cell. These should be formed from two orbitals per V
using V-V hopping as described in the tight binding discussion above.
The light bands, which provide the carriers, should have width 2 eV,
$U/W \alt 1$ and containing $\approx$ 0.5 electrons per V.
In addition, there should be a localized
set of atomic-like orbitals (one per V) with occupation $1 e$ and bare band
width $\approx$ 0.5 eV. Additionally, the off-diagonal Coulomb repulsion
coupling these two groups of states should be comparable, i.e. $\approx$
2 eV. LiV$_2$O$_4$ can be made quite cleanly and is well suited to
detailed experimental investigation, and this may make it an especially
interesting system for experimental and theoretical study.

\acknowledgements

Computations were performed using facilities of the DoD HPCMO ASC and
NAVO Centers.
Work at the Naval Research Laboratory is supported by the
Office of the Naval Research.

\end{multicols}

\begin{references}
\bibitem{lapwnote}
Highly converged calculations were performed with two separate codes,
WIEN-97, (ref. \onlinecite{blahalapw})
and an independent general potential LAPW code, (ref.  \onlinecite{weilapw})
using different choices of sphere radii,
zone integration and other parameters. No significant differences
were found in parallel calculations.
\end{references}
\end{document}